\documentclass[aps,prb,superscriptaddress,showpacs,floatfix,twocolumn]{revtex4}
\usepackage{amsmath,amsfonts}
\usepackage{times,mathptm}
\usepackage{color}
\usepackage{epsfig}
\usepackage{dcolumn}
\begin{document}
\title{Substrate effects on surface magnetism of Fe/W(110) from first principles}
\author{Torsten Andersen}
\affiliation{Department of Physics, Kaiserslautern University of Technology,
 Box 3049, D--67653 Kaiserslautern, Germany}
\author{Wolfgang H{\"u}bner}
\affiliation{Department of Physics, Kaiserslautern University of Technology,
 Box 3049, D--67653 Kaiserslautern, Germany}
\date{\today: Revised}
\begin{abstract}
 Surface magnetic properties of the pseudomorphic Fe(110) monolayer on
 a W(110) substrate are investigated from first principles as a
 function of the substrate thickness (up to 8 layers). Analyzing the
 magnetocrystalline anisotropy energies, we find stable (with respect
 to the number of substrate layers) in-plane easy and hard axes of
 magnetization along the [1$\bar{1}$0]- and [001]-directions,
 respectively, reaching a value in good agreement with experiment for
 thick substrates.  Additionally, the changes to the magnetic spin
 moments and the density of the Fe $d$-states are analyzed with
 respect to the number of substrate layers as well as with respect to
 the direction of magnetization. With respect to the number of W(110)
 substrate layers beneath the Fe(110) surface, we find that the first
 four substrate layers have a large influence on the electronic and
 magnetic properties of the surface. Beyond the 4th layer, the
 substrate has only marginal influence on the surface properties.
\end{abstract}
\maketitle
\section{Introduction}
The pseudomorphic monolayer of Fe grown on a W(110) surface is very
interesting from the point of view of magnetism. In studies of the
surface magneto-crystalline anisotropy, the Fe monolayer on top of a W
substrate has become the system of choice, since (i) the growth of the
first Fe monolayer is pseudomorphic, (ii) the W substrate has a large
spin-orbit coupling, and (iii) the interface anisotropy is the
strongest ever observed. This makes the Fe monolayer on a W substrate
a good candidate for an {\em ab initio} benchmark investigation of how
the properties of the magneto-crystalline anisotropyare influenced by
the substrate.

Depending on the coverage and the surface orientation, Fe films on a W
substrate show very different magnetic properties. At submonolayer
coverages on the (110) surface one observes the formation of separate
islands that are nonmagnetic up to a coverage of 58-60\%, beyond which
they become ferromagnetic islands,\cite{c:1} which may even show an
out-of-plane magnetization until they approach the full-monolayer
coverage.\cite{c:2} A single monolayer (ML) of Fe can be grown
pseudomorphically on top of a W(110) surface.\cite{c:3} Experiments
performed on the pseudomorphic Fe
monolayer\cite{c:9,c:4,c:5,c:6,c:7,c:8} have shown that\cite{c:5}
``the prominent magnetic feature of Fe(110) films on W(110) is a
strong in-plane magnetic surface anisotropy, with an easy axis
[1$\bar{1}$0] at right angles to the bulk easy axis [001].'' At
coverages of about 1.5 ML, double-layer patches (sesquilayers) form
with an out-of-plane magnetic easy axis,\cite{c:10} and an
antiferromagnetic order.\cite{c:11,c:12} A pseudomorphic full second
ML has to our knowledge not been grown experimentally, but theory
predicts it would also have an in-plane easy axis in the
$1\bar{1}0$-direction.\cite{c:13} Attempts of using annealing as a
means of making the grown Fe layers full monolayers beyond the first
monolayer leads to a reorganization of the Fe in an Fe bulk-like
lattice.\cite{c:14} For continous layers of Fe, the easy axis of
magnetization eventually changes to the in-plane [001]-direction it
has in bulk Fe,\cite{c:15,c:16,c:17,c:18} the critical thickness of
the Fe film being reported in the range between 80 and 95 \AA\ (95
\AA\ reported by Ref.~\onlinecite{c:16}, and 80--86 \AA\ reported by
Ref.~\onlinecite{c:18}).

In passing, we should stress that in contrast to the ferromagnetic Fe
monolayer on the W(110) surface stand results of recent studies
(experimental\cite{c:19,c:20} and {\em ab
initio})\cite{c:21,c:20,c:22} of Fe films on the (001) surface of W,
which indicate that the first ML of Fe is antiferromagnetic when grown
on the (001) surface. Thus, the magnetic properties of the Fe
monolayer is very sensitive to the surface orientation of the W
substrate, and one can not make general conclusions about the magnetic
properties of other surfaces on the basis of investigations of only
one.

In the present work, we study the substrate effects on the magnetic
properties of the pseudomorphic ML of Fe(110) on W(110) from first
principles. Existing {\em ab initio} calculations performed within the
full-potential linearized augmented plane-wave (FLAPW)
method,\cite{c:23,c:24,c:13} the full-potential linearized muffin-tin
orbital (FP-LMTO) method,\cite{c:25} as well as pseudo-potential
calculations\cite{c:21} confirm the above-mentioned experimental
ferromagnetic result with an easy axis along the
[1$\bar{1}$0]-direction. At present, there is no measurable
lattice-relaxation effect of the magnetization direction,\cite{c:26}
and the starting point for the calculations presented in this work is
the optimized surface of Ref.~\onlinecite{c:23}.  The main results of
our work are magnetocrystalline anisotropy energies that are converged
{\em ab initio} as a function of the substrate thickness, along with
magnetic spin- and orbital moments that consistently confirm the
breaking of Hund's 3rd rule for the W substrate. The quality of the
results is illustrated by explicitly showing the convergence, along
with an estimate of the accuracy.

The paper is organized as follows: in Sec.~\ref{sec:II} we describe
and discuss the method that we use, in particular with respect to the
level of accuracy necessary for the calculation of stable
magnetocrystalline anisotropy energies, and in Sec.~\ref{sec:III} we
present and discuss the results of our calculations, where we focus
the attention on the magnetic and electronic properties of the
surface, and how the substrate influences these properties. Finally,
in Sec.~\ref{sec:IV}, we conclude.

\section{Method}\label{sec:II}
Since we have the flexibility that a theoretical calculation provides,
we do not need to limit ourselves to the experimentally
possible. Thus, we are able to take an approach where we begin with a
bilayer of Fe(110) and W(110), successively adding W substrate layers
underneath in order to investigate how the the sub-surface layers
influences the magnetic and electronic properties at the surface.

In order to facilitate the calculation of magnetocrystalline
anisotropy energies and how they change with the number of substrate
layers, one has to perform the calculations with a highly accurate and
mature {\em ab initio} computer code. The demands on accuracy imply
that a full-potential code is necessary. Since all {\em ab initio}
methods employs an ``educated guess'' for the initial solution, the
code {\em must} be self-consistently reaching a stable solution to the
eigenvalue problem of the Schr{\"o}dinger equation, e.g., by use of
the Rayleigh-Ritz\cite{c:27,c:28} variational approach. In addition to
this, the spin-orbit interaction must be included in the
self-consistent solution in order to produce magnetic properties that
are accurate enough. The {\em ab initio} method of choice for our
calculations is the full-potential (partially linearized) augmented
plane-wave [(L)APW+lo] with local orbitals method\cite{c:29} as
implemented in the {\sc Wien2k} computer code.\cite{c:30}

Within the {\sc Wien2k} code, one constructs a supercell perpendicular
to the surface, ensuring that enough vacuum is inserted between the
surface and the border of the unit cell (in our case the vacuum
amounts to at least 16 interlayer distances of the W(110) substrate).
In order to keep the environment of our surface calculation stable, we
let the Fe monolayer sit in the middle of the unit cell of constant
size, adding the W substrate layers under it. This choice, of course,
limits our possibility for continously adding W substrate layers, but
not beyond the practical limits of the code, as we shall see later.

In a study where we begin from the surface layer we expect to see
rather big changes to the electronic and magnetic properties as we add
substrate layers to the system, due to the similarities to the
properties of a quantum well. As we keep adding substrate layers we
expect the properties to stabilize in such a manner that the system
starts behaving like a surface layer on a bulk-like substrate. When
this point has been reached, adding another substrate layer should not
change the electronic and magnetic properties at the Fe surface layer
much.

With our demands on the accuracy, it is important that the surface is
structurally relaxed. How a single monolayer of Fe(110) relaxes on top
of a W(110) substrate has been studied by Qian and H{\"u}bner
(Ref.~\onlinecite{c:23}). Since their results are very close to
experiment,\cite{c:4,c:5,c:1,c:6,c:11,c:31,c:7,c:8} and since there is
no experimentally measurable structural effect due to changes in the
direction of magnetization,\cite{c:26} we have in the present work
chosen to adopt the optimized surface structure of
Ref.~\onlinecite{c:23} as the basis for our calculations. We have,
however, in order to better describe the influence of the substrate on
the surface magnetism, chosen to use a nonsymmetric layered structure
instead of the symmetric one of Ref.~\onlinecite{c:23}. Thus, the
structure is using the optimized values of the interlayer distances
from the surface side as determined in Ref.~\onlinecite{c:23} [the
Fe-W1 distance is contracted by 12.9\%\ and the W1-W2 distance by
0.1\% compared to bulk W(110) interlayer distances (W1 and W2 are the
first and second W layers under the Fe layer, respectively)], but has
bulk-spaced W layers underneath. The Fe layer has an in-plane lattice
misfit of about 10\%\ compared to a pure Fe(110) surface.

In practical {\em ab initio} calculations within density-functional
theory, one has to make a choice of the so-called exchange-correlation
potential, since it is not known as an exact quantity. Our
calculations within the {\sc Wien2k} computer code are performed using
the [generalized gradient approximation (GGA)] exchange-correlation
potential of Perdew, Burke, and Ernzerhof (Ref.~\onlinecite{c:32}). In
order to facilitate the calculation of magnetic anisotropy energies
(MAEs), the variational parameter, i.e., the total energy of the {\em
magnetic} configuration,\cite{c:32a} was converged to an accuracy that
in most cases is better than 10 $\mu$Ry (for a detailed list of
relevant accuracies, consult the results section, in particular
Tab.~\ref{t:1}). Also, the fluctuations in the charge density within
the unit cell has been converged to less than 10$^{-4}$
e/Bohr$^3$. Since the abovementioned two convergence criteria do not
automatically ensure that the eigenstates and eigenfunctions are
sufficiently accurate, one has in addition to make sure that also the
the calculation converges with respect to the parameters that control
the accuracy of the calculations of the {\sc Wien2k} code [in
particular the (kinetic energy) cutoff value of the otherwise infinite
plane-wave basis, and the sampling in {\bf k}-space]. As we shall
illustrate below, these parameters were adjusted (increased) until the
magnetic anisotropy energies for an increasing number of W layers
stabilized.

The parameter used to control the kinetic energy cutoff of the
plane-wave basis depends on the muffin-tin radius used for the atomic
part of the (L)APW basis set as follows:
\begin{equation}
T_c=R_{\sf{MT}}k_{\sf{max}},\label{e:1}
\end{equation}
where $T_c$ is the cutoff parameter, $R_{\sf{MT}}$ is the muffin-tin
radius (2.35 Bohr in our calculations), and $k_{\sf{max}}^2$
corresponds to the plane-wave cutoff (in Ry) in pseudopotential
calculations, e.g., the value $T_c=9$ gives a $k_{\sf{max}}^2=199.56$
eV. Thus, $T_c$ determines the matrix size of the eigenvalue problem,
and higher values of $T_c$ potentially decreases the accuracy of the
resulting eigenfunctions and -energies.

\section{Results and discussion}\label{sec:III}
When one begins with an Fe/W(110) bilayer and thereafter add W(110)
substrate layers (on the W side of the bilayer), one expects that as
long as the number of W layers (called $N_W$ in the following) is
sufficiently small, the system behaves in a quantum-well-like fashion,
i.e., the changes to properties such as the magnetocrystalline
anisotropy energy or the density of states will be large each time a
substrate layer is added. At some critical value of $N_W$, one would
expect that the substrate starts behaving like a bulk substrate and,
thus, that the Fe layer becomes more surface-like in its
properties. Below, we discuss how the magnetocrystalline anisotropy
energies and the electronic structure change as a function of $N_W$.

\begin{table*}[t]
\caption{(Color online) Evolution of the magnetocrystalline anisotropy
energy of the Fe/W(110) surface as a function of (i) the number of W
layers beneath the pseudomorphic Fe monolayer included in the
calculation, $N_W$, (ii) the kinetic energy cutoff of the plane-wave
basis, $T_c=R_{\sf{MT}}k_{\sf{max}}$ [below, $T_c=7$ corresponds to
$k_{\sf{max}}^2=120.72$ eV, $T_c=8$ to $k_{\sf{max}}^2=157.68$ eV, and
$T_c=9$ to $k_{\sf{max}}^2=199.56$ eV], and (iii) the number of {\bf
k}-points in the full Brillouin zone, \#{\bf k}. The leftmost half of
the table shows how the total energy of the surface magnetized along
the easy axis, {\bf M}$\|$[1$\bar{\mbox{1}}$0] evolve as a function of
$T_c$, \#{\bf k}, and $N_W$. The first column is the number of W
layers, the second column is the total energy at
$\{T_c,\#{\bf{k}}\}=\{7,441\}$, columns 2--6 show the evolution of
this total energy as the accuracy of the calculation increases (as the
difference in total energy with respect to the previous calculation),
and in column 7 is shown the total energy of the converged calculation
along with the uncertainty estimate, as given by Eq.~(\ref{e:2}). In
columns 8--12 is shown how the MAE of the perpendicular magnetization
direction ({\bf M}$\|$[110]) changes during convergence, where in
column 12 the uncertainty estimate has been added. Column 13 lists the
converged MAEs for {\bf M}$\|$[001] with their respective uncertainty
estimates (for the uncertainties of {\bf M}$\|$[001] relative to {\bf
M}$\|$[110], see the text). Each and every number in this table have
been determined self-consistently.\label{t:1}}
\begin{ruledtabular}
\begin{tabular}{c|rrrrrr|r||rrrrr|r}
 &\multicolumn{7}{c||}{Total energy for {\bf M}$\|$[1$\bar{\mbox{1}}$0]}
 &\multicolumn{6}{c}{MAE with respect to {\bf M}$\|$[1$\bar{\mbox{1}}$0] (meV)} \\
 &\multicolumn{2}{c}{(MeV)} & \multicolumn{4}{c|}{evolution (meV)} 
&\multicolumn{1}{c||}{Final (meV)} & \multicolumn{5}{c|}{{\bf M}$\|$[110]} & {{\bf M}$\|$[001]}  \\
\cline{2-14}
 & $T_c$: & \multicolumn{1}{c}{7} & \multicolumn{1}{c}{8} & \multicolumn{1}{c}{8} 
& \multicolumn{1}{c}{9} & \multicolumn{1}{c|}{9} & \multicolumn{1}{c||}{9} 
& \multicolumn{1}{c}{7} & \multicolumn{1}{c}{8} & \multicolumn{1}{c}{8} & \multicolumn{1}{c}{9} 
& \multicolumn{1}{c|}{9} 
& \multicolumn{1}{c}{9} \\
$N_W$ & \#{\bf k}:& \multicolumn{1}{c}{441} & \multicolumn{1}{c}{441} 
& \multicolumn{1}{c}{961} & \multicolumn{1}{c}{961} & \multicolumn{1}{c|}{2025} 
& \multicolumn{1}{c||}{2025} &\multicolumn{1}{c}{441} & \multicolumn{1}{c}{441} 
& \multicolumn{1}{c}{961} & \multicolumn{1}{c}{961} 
& \multicolumn{1}{c|}{2025} 
& \multicolumn{1}{c}{2025} \\
\hline
1 & \multicolumn{2}{r}{-0.474538} &  -574 &  0.75 & -157 &  0.18 
&{\color{red}-474539186.379$\pm$0.034}
& 0.80 & -0.03 & -0.80 & 0.29 & {\color{red}0.11$\pm$0.08} 
& {\color{red}1.77$\pm$0.12} 
\\
2 & \multicolumn{2}{r}{-0.914445} &  -917 &  0.46 & -257 &  0.11 
&{\color{red}-914445985.433$\pm$0.068}
& 4.53 &  4.61 &  3.81 & 4.63 & {\color{red}4.75$\pm$0.10} 
& {\color{red}6.00$\pm$0.11} 
\\
3 & \multicolumn{2}{r}{-1.354351} & -1257 & -0.44 & -354 &  0.53 
&{\color{red}-1354352832.665$\pm$0.034}
& 2.57 &  2.79 &  2.12 & 3.29 & {\color{red}3.16$\pm$0.16} 
& {\color{red}3.66$\pm$0.11} 
\\
4 & \multicolumn{2}{r}{-1.794257} & -1595 & -1.13 & -448 & -0.33 
&{\color{red}-1794259724.579$\pm$0.061}
& 2.45 &  3.29 &  2.76 & 2.68 & {\color{red}2.34$\pm$0.15} 
& {\color{red}2.82$\pm$0.19} 
\\
5 & \multicolumn{2}{r}{-2.234164} & -1931 & -0.07 & -546 & -0.73 
&{\color{red}-2234166567.988$\pm$0.041}
& 3.61 &  2.34 &  0.11 & 1.97 & {\color{red}2.08$\pm$0.15} 
& {\color{red}2.57$\pm$0.21} 
\\
6 & \multicolumn{2}{r}{-2.674070} & -2263 & -3.20 & -638 & -0.41 
&{\color{red}-2674073400.852$\pm$0.054}
& 2.04 &  2.03 &  0.98 & 1.97 & {\color{red}2.60$\pm$0.18} 
& {\color{red}2.79$\pm$0.14} 
\\
7 & \multicolumn{2}{r}{-3.113976} & -2592 & -1.85 & -738 & -0.80 
&{\color{red}-3113980210.383$\pm$0.136}
& 2.80 &  1.67 &  0.19 & 2.07 & {\color{red}2.72$\pm$0.31} 
& {\color{red}3.01$\pm$0.38} 
\\
8 & \multicolumn{2}{r}{-3.553883} & -2921 & -5.07 & -823 &  0.24 
&{\color{red}-3553887008.444$\pm$0.258}
& 3.70 &  2.27 & -0.03 & 2.29 & {\color{red}2.26$\pm$0.53} 
& {\color{red}2.99$\pm$0.41} 
\\
\end{tabular}
\end{ruledtabular}
\end{table*}

\subsection{Magnetocrystalline anisotropy energies}
Searching for magnetocrystalline anisotropy energies on the basis of
{\em ab initio} methods that minimizes the total energy, one is
looking for energy differences in the meV-range which has to be
calculated from differences between total energies in the
MeV-range. In order to provide us with two significant digits in the
meV-range it is required that the computed total energies are
numerically accurate to (within the limits of the physical model) the
11th decimal. Todays standard arithmetic precision (64-bit numbers
with a 52-bit mantissa) provides 16 significant digits, and thus
already a sum over 1000 {\bf k}-points reduces precision to 13
significant digits at best. Carefully written self-consistent
full-potential methods may limit the further loss of precision caused
by sampling, but the amount of {\bf k}-points is limited if one wants
to obtain meaningful results.

In order to be able to draw any conclusions about the substrate
effects and the properties of the surface for a fixed geometry, the
calculation of the MAEs must first converge with respect to, in
particular, the kinetic energy cutoff of the plane-wave basis, as well
as the sampling in {\bf k}-space.\cite{c:32b} Since MAEs are
calculated here as differences of ``total energies'', this requirement
automatically applies to the ``total energies'' (even though the
``total energies'' are not necessarily physically meaningful
quantities themselves). Thus, the evolution of the magnetocrystalline
anisotropy energies of the Fe/W(110) surface is described in
Tab.~\ref{t:1} as a function of (i) the number of W layers added
underneath the pseudomorphic Fe monolayer, (ii) the kinetic energy
cutoff of the plane-wave basis, as defined by Eq.~\ref{e:1}, and (iii)
the number of sampling points ({\bf k}-points) used in the full
Brillouin zone (FBZ), here called \#{\bf k} in a short notation. In
order to make the numbers directly comparable and to avoid pure {\bf
k}-point effects, calculations on the same level of accuracy have been
performed with the {\em same} {\bf k}-points (in the FBZ), and the
volume of the ``unit cell'' has been kept constant. Since MAEs are
differences in ``total energies'', Tab.~\ref{t:1} consists of two main
blocks, namely (left) the evolution of the total energy of the ground
state, which is the system that is magnetized along the {\em easy}
magnetic axis, here {\bf M}$\|$[1$\bar{\mbox{1}}$0] in agreement with
experiment (see
Refs.~\onlinecite{c:4,c:5,c:1,c:6,c:11,c:31,c:33,c:7,c:8}) and theory
(see Ref.~\onlinecite{c:23}); and (right) the evolution in the MAEs
with respect to the easy axis for the out-of-plane direction, {\bf
M}$\|$[110]. The rightmost column shows the converged MAEs for the
in-plane direction that is perpendicular to the easy axis, i.e., the
{\em hard} axis {\bf M}$\|$[001].

The far left column lists the number of W layers added underneath the
pseudomorphic Fe surface layer.  Columns 2--7 show results for the
total energy of the systems with magnetization along the easy
axis. Column 2 shows the total energy in MeV of the cheapest
calculation (lowest value of $T_c$ and lowest number of {\bf
k}-points) performed in this study. Column 3 shows the difference in
the total energy (with respect to the results in column 2) when one
increases the kinetic energy cutoff. One observes that the total
energy increases with about 340 meV per W atom. In column 4, the
density of the sampling in {\bf k}-space is increased, with changes in
the total energy compared to the results in column 3 of less than a
meV per atom. Increasing $T_c$ again in column 5, the total energy
changes at most 100 meV per W atom, and the calculation has become so
stable that a further increase in the sampling in {\bf k}-space
(column 6) gives rise to changes in the total energy of less than 1
meV. Increasing $T_c$ beyond the value of 9 introduces ghost bands,
and self-consistent minimization of the total energy is not possible.
With the small changes in the total energy occuring in column 6,
however, together with the stability seen in the MAEs in the right
half of the table, leads us to conclude that the calculation has
converged with respect to $T_c$ and \#{\bf k}. Thus, in column 7 we
have listed the total energies in meV for the calculation from column
6 to the $\mu$eV-level, together with an estimate of the uncertainty
obtained in the self-consistent minimization procedure. This
uncertainty is calculated from the total energies of the last three
cycles of the self-consistent minimization of the total energy. Let
$n$ refer to the last cycle, then the uncertainty $\varepsilon(n)$ on
the total energy $V(n)$ in the last cycle is calculated as:
\begin{equation}
 \varepsilon(n)=\left[\left|V(n)-V(n-1)\right|
 +\left|V(n)-V(n-2)\right|\right]/2 .
\label{e:2}
\end{equation}
The smallest uncertainty achieved in the present calculation is 2.5
$\mu$Ry (34 $\mu$eV). We notice in passing that at 7 and 8 substrate
layers, the uncertainties become larger. Beyond 8 substrate layers,
the calculations become less stable, i.e., the required level of accuracy
in order to calculate well converged MAEs could not be reached.

In the right half of Tab.~\ref{t:1} are listed the MAEs that come out,
when the calculations in columns 2--7 are repeated for {\bf
M}$\|$[110] and the results for the easy-axis calculation are
subtracted. Hence, since all ``total energies'' are negative, positive
numbers indicate that the absolute value of the total energy is
smaller than that of the easy-axis calculation, and thus we can
conclude that the axis is ``harder''. Similarly, negative values
indicate that the axis is ``easier'', and one notices immediately that
in the two right-most calculations (columns 11 and 12) there are no
negative values (for $T_c=9$). Column 13 lists the MAEs for {\bf
M}$\|$[001]. Here we have left out the evolution, since it shows
behaviour similar to the MAEs for {\bf M}$\|$[110].  It is important
to underscore the fact that each of the numbers that appear in
Tab.~\ref{t:1} is a result of an individual self-consistent
minimization of the total energy and the fluctuations in the charge
density for the specific configuration. In Tab.~\ref{t:1}, three
columns are in red. The numbers in these columns are the final,
converged, results, and are listed with their individual level of
uncertainty. Naturally, the uncertainties of the total energy for the
easy axis are small compared to those of the MAEs. The uncertainties
listed for the MAEs are calculated as the sums of the uncertainties
for the total energies of the two calculations involved in getting a
single MAE. One should bear in mind that the uncertainties between the
results for {\bf M}$\|$[110] and {\bf M}$\|$[001] are {\em not} simply
an addition of the uncertainties listed in columns 12 and 13 of
Tab.~\ref{t:1}, since the uncertainties listed are on the MAEs, {\em
not} the total energy, and thus with respect to {\bf
M}$\|$[1$\bar{\mbox{1}}$0]. The relative uncertainties between the
results for {\bf M}$\|$[110] and {\bf M}$\|$[001] are (in meV):
$\pm$0.20 for 1 W layer, $\pm$0.07 for 2 W layers, $\pm$0.20 for 3 W
layers, $\pm$0.31 for 4 W layers, $\pm$0.19 for 5 W layers, $\pm$0.16
for 6 W layers, $\pm$0.39 for 7 W layers, and $\pm$0.37 for 8 W
layers. With these values in mind, it is safe to conclude that the
magnetic {\em hard} axis is in the [001]-direction.

\begin{figure}[tb]
\centerline{\epsfig{file=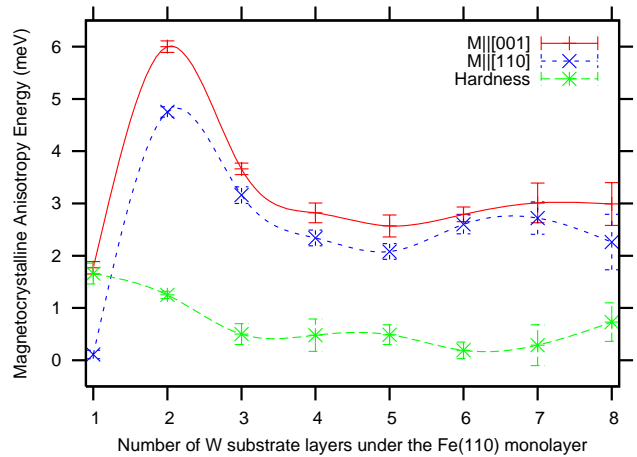,width=8.5cm}}
\caption{(Color online) Magnetic anisotropy energies relative to the
easy axis ({\bf M}$\|[1\bar{1}0]$) are shown as a function of the
number of W substrate layers underneath the Fe(110) surface
layer. Data are taken from Tab.~\ref{t:1} and the text, and the
uncertainties have been included as vertical error bars in the
plot. The red (full line) and blue (short-dashed line) curves show the
MAEs of the [001]- and [110]-directions, respectively. The green
(long-dashed line) curve shows the energy difference between samples
magnetized along the [001] and [110] directions, which also can be
called the degree of hardness of the hard axis (see text). To guide
the eye, cubic splines have been used to connect the data
points.\label{f:1}}
\end{figure}

In order to visualize the influence of the substrate on the
magnetocrytalline anisotropy energies, we have in Fig.~\ref{f:1}
plotted the MAEs relative to the easy axis ({\bf M}$\|[1\bar{1}0]$)
with respect to the number of W layers in the substrate under the
Fe(110) monolayer, along with their respective uncertainties
(indicated by a vertical bar at each data point). In order to guide
the eye, cubic splines have been included to connect the data
points. The red (upper) line shows the evolution of the MAE of the
in-plane hard [001] axis, while the blue line shows how the MAE of the
direction perpendicular to the surface [110] evolves when adding
substrate layers. The green (bottom) line shows the MAE between the
hard and the perpendicular axes, and can thus be interpreted as the
``hardness'' of the hard axis. As long as it is positive, the hard
axis is along the in-plane [001]-direction, whereas if it became
negative the hard axis would shift to being out-of-plane.

The trend in the values of the MAEs is clear. At one layer of W, the
MAEs are small, but at two layers of W one sees a dramatic increase,
and the MAEs reaches their maximum values. Adding a 3rd layer of W,
the MAEs drop again, and from the 4th layer and on they get almost
constant, in agreement with Ref.~\onlinecite{c:25}. The value at the
``bulk'' end (7--8 substrate layers) is, with its 3 meV quite close to
the experimental value of 4.2 meV obtained recently by Pratzer,
Elmers, {\em et al.} (Ref.~\onlinecite{c:33}) for Fe monolayer stripes
on W(110). That the values differ by about 1 meV could have its origin
in the fact that the experiment is made on stripes. Since the stripes
have lower symmetry than a perfect monolayer we would expect the MAE
to be slightly higher for the stripes. Older
experiments\cite{c:12,c:17,c:33a,c:33b} give values of the MAEs in the
range from about 0.11 meV to about 6.5 meV, depending on the setup,
temperature, and film thickness.

Table \ref{t:1} and Fig.~\ref{f:1} reveal a magnetic hard axis along
the $[001]$-direction. That the absolute values\cite{c:34} differ
from those of Ref.~\onlinecite{c:13} can be attributed to the fact
that we are using asymmetric structures (with an Fe layer only on one
side of the slab) in the present work, whereas Ref.~\onlinecite{c:13}
used symmetric structures (with an Fe layer on both sides of the
slab).

\begin{table}[t]
\caption{Magnetic anisotropy energies for different magnetization
directions in 1 ML of Fe on top of 4 ML of W(110). The anisotropy
energies are with respect to the value of the total energy for {\bf
M}$\|[1\bar{1}0]$. Here, $T_c=9$ and \#{\bf k}=2025, and the
uncertainties are calculated as in Tab.~\ref{t:1} by use of
Eq.~(\ref{e:2}). The first four rows are in-plane and the last four
rows out-of-plane (see text).\label{t:2}}
\begin{ruledtabular}
\begin{tabular}{cc}
Magnetization direction & MAE (meV) \\ \hline
$5^\circ$ from $[1\bar{1}0]$ towards $[001]$ & 0.15$\pm$0.09 \\ 
$[1\bar{1}1]$ & 0.83$\pm$0.17 \\
$5^\circ$ from $[001]$ towards $[1\bar{1}0]$ & 2.82$\pm$0.09 \\ 
$[001]$ & 2.82$\pm$0.19\\ \hline
$5^\circ$ from $[001]$ towards $[110]$ & 2.53$\pm$0.22 \\ 
$5^\circ$ from $[1\bar{1}0]$ towards $[110]$ & 2.31$\pm$0.21 \\ 
$5^\circ$ from $[1\bar{1}1]$ towards $[110]$ & 2.59$\pm$0.20 \\ 
$[110]$ & 2.34$\pm$ 0.15\\
\end{tabular}
\end{ruledtabular}
\end{table}

\begin{figure*}[tb]
\centerline{\epsfig{file=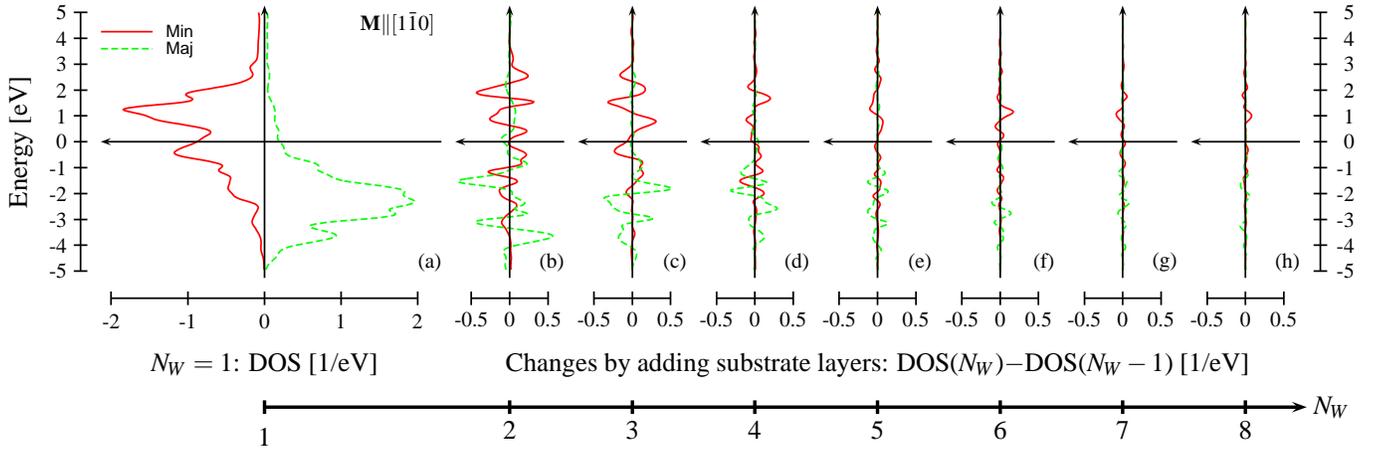,width=18cm}}
\caption{(Color online) Evolution in the density of d-states (d-DOS)
for an Fe(110) monolayer magnetized along the easy axis ({\bf
M}$\|[1\bar{1}0]$) when up to 8 W(110) substrate layers are grown
underneath it. The plot to the left (a) shows the density of states
for the d-electrons of the Fe atom for an Fe monolayer with a single W
substrate layer underneath. The d-DOS for the minority spin has been
plotted with reverse sign for ease of understanding. The plots
(b)--(h) shows the {\em changes} in the d-DOS of the Fe surface layer
when additional W substrate layers are added underneath. In order to
guide the eye, the scales are kept in constant proportion in all
plots. On the energy scales, the Fermi energy is taken as the
reference point ($E_F=0$). In all plots, red (full) lines are the
results for the minority spin channel and green (dashed) lines are the
results for the majority spin channel. In order to assure a good view
of the plots we have moved the scales to the border of the
figure. \label{f:2}}
\end{figure*}

\begin{figure}[tb]
\centerline{\epsfig{file=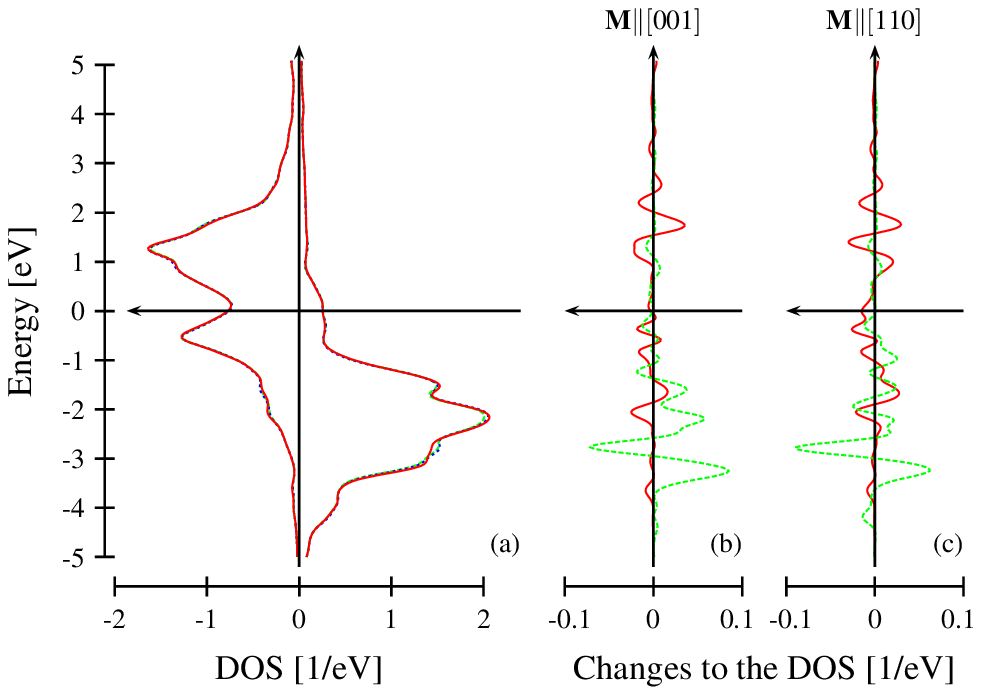,width=8.5cm}}
\caption{(Color online) The density of d-electrons of the Fe atom
(d-DOS) is shown for the calculation that has converged with respect
to the number of substrate layers, i.e., for $N_W=4$. Results are
shown in the combined plot (a) for the easy (red, full line), hard
[green, long-dashed line), and perpendicular (blue, short-dashed line)
directions of magnetization, and (b)--(c) show the differences in the
Fe d-DOS between (b) the easy and hard axes and between (c) the easy
and perpendicular axes, respectively. In (b) and (c), the differences
in the minority spin channel are plotted in red (full lines), and the
differences in the majority spin channel in green (dashed lines). For
ease of understanding, the d-DOS of the minority spin has been plotted
with reverse sign, as in Fig.~\ref{f:2}. Again, the zero on the energy
scale corresponds to the Fermi energy and, again, we have put the
scales to the border of the figure.\label{f:3}}
\end{figure}

\subsection{Magnetic properties of the electronic structure}
In order to verify that the easy axis is really along one of the main
crystallographic axes (and along the $[1\bar{1}0]$-direction) we have
performed a number of calculations of the total energy where the
magnetization direction has been shifted by $5^\circ$ away from the
main crystallographic axes (in various directions, as indicated in
Tab.~\ref{t:2}). These calculations have been performed for four
substrate layers only. Thus, we have in Tab.~\ref{t:2} printed the
MAEs with respect to the magnetic easy axis for $T_c=9$ and \#{\bf
k}$=2025$. In Tab.~\ref{t:2}, the first four lines show the
development of the MAE as we change the magnetization from the easy
axis to the hard axis in the surface plane, the next three lines show
how the MAE changes as we go away from the surface, and the last line
shows the value perpendicular to the surface. The values along the
hard and perpendicular axes are taken from Tab.~\ref{t:1}. It is
important to stress that in order for the results in Tab.~\ref{t:2} to
be trustworthy (since the MAEs are calculated from total energies),
one needs to make the calculations identical in the sense that {\em
only} the direction of magnetization is allowed to change. The
critical quantity to keep identical here is the {\bf k}-space
sampling-mesh (total energies are {\bf k}-dependent, as is evident
from Tab.~\ref{t:1}). The results in Tab.~\ref{t:2} are therefore all
obtained with the exact same {\bf k}-points (in the Full Brillouin Zone),
regardless of the changes in symmetry.

Of the first four lines in Tab.~\ref{t:2}, the first line shows the
result when the magnetization direction is shifted by $5^\circ$ in the
direction of the hard axis. When the magnetization is shifted further
in the direction of the hard axis, one passes the crystallographic
$[1\bar{1}1]$-direction (second line of Tab.~\ref{t:2}) that could be
a natural place to look for an easy or hard axis of magnetization. As
we can see, it is neither the easy- nor the hard axis. Close to the
hard axis (c), the anisotropy at $5^\circ$ away from it has a value
that is very close to (equal to, in fact, within our precision) that
of the hard axis. The value for the hard axis MAE is shown for
reference in the fourth line. Going $5^\circ$ out of the surface from
the three axes $[1\bar{1}0]$, $[1\bar{1}1]$, and $[001]$, we again see
values between those of the easy and hard axes. In fact, the change in
the MAE by going $5^\circ$ out of the surface plane from the easy axis
is quite strong. Thus, we conclude from Tab.~\ref{t:2} that evidence
is strong enough for to confirm that the easy axis is along the
(in-plane) $[1\bar{1}0]$-direction, in agreement with previous
experimental\cite{c:4,c:5,c:1,c:6,c:11,c:31,c:7,c:8} and
theoretical\cite{c:13,c:25} findings.

\begin{table*}[bt]
\caption{Magnetic spin ($\mu_S$) and orbital ($\mu_L$) moments per
atom (in Bohr magnetons, $\mu_B$) are shown for the pseudomorphic Fe
surface layer as well as the first W substrate layer (W$_1$) as a
function of the number of substrate layers ($N_W$) when the
magnetization is along the in-plane easy axis (columns 2--5), the
in-plane hard axis (columns 6--9), and perpendicular to the surface
(columns 10--13).\label{t:3}}
\begin{ruledtabular}
\begin{tabular}{ccccccccccccc}
 &\multicolumn{4}{c}{Easy axis: {\bf M}$\|[1\bar{1}0]$}
 &\multicolumn{4}{c}{Hard axis: {\bf M}$\|[001]$}
 &\multicolumn{4}{c}{Perpendicular: {\bf M}$\|[110]$} \\
 &\multicolumn{2}{c}{Surface: Fe} & \multicolumn{2}{c}{Substrate: W$_1$}
 &\multicolumn{2}{c}{Surface: Fe} & \multicolumn{2}{c}{Substrate: W$_1$}
 &\multicolumn{2}{c}{Surface: Fe} & \multicolumn{2}{c}{Substrate: W$_1$}\\
$N_W$&$\mu_S$& $\mu_L$  & $\mu_S$&$\mu_L$ &$\mu_S$&$\mu_L$   &$\mu_S$ &$\mu_L$ &$\mu_S$& $\mu_L$  & $\mu_S$&$\mu_L$ \\
\hline
  1  & 2.562 & -0.00023 & -0.135 &  5.792 & 2.561 &  0.00005 & -0.136 & -2.184 & 2.560 &  0.00007 & -0.135 & -0.584 \\ 
  2  & 2.574 & -0.00010 & -0.100 &  2.517 & 2.574 &  0.00009 & -0.099 & -3.449 & 2.570 &  0.00007 & -0.102 & -1.359 \\ 
  3  & 2.616 & -0.00006 & -0.089 & -0.021 & 2.613 &  0.00005 & -0.091 &  0.021 & 2.609 &  0.00005 & -0.094 &  0.024 \\ 
  4  & 2.561 & -0.00006 & -0.104 & -0.025 & 2.557 &  0.00005 & -0.105 &  0.024 & 2.559 &  0.00005 & -0.104 &  0.028 \\ 
  5  & 2.567 & -0.00006 & -0.098 & -0.024 & 2.563 &  0.00004 & -0.100 &  0.023 & 2.565 &  0.00005 & -0.099 &  0.026 \\ 
  6  & 2.568 & -0.00005 & -0.101 & -0.024 & 2.567 &  0.00004 & -0.101 &  0.023 & 2.567 &  0.00005 & -0.100 &  0.027 \\ 
  7  & 2.559 & -0.00005 & -0.104 & -0.025 & 2.558 &  0.00004 & -0.104 &  0.024 & 2.558 &  0.00005 & -0.104 &  0.028 \\ 
  8  & 2.565 & -0.00005 & -0.103 & -0.025 & 2.563 &  0.00004 & -0.103 &  0.024 & 2.562 &  0.00005 & -0.103 &  0.028
\end{tabular}
\end{ruledtabular}
\end{table*}

Tab.~\ref{t:3} lists the magnetic spin- ($\mu_S$) and orbital
($\mu_L$) moments per atom for the three different magnetization
directions we consider. First, we notice in general that the magnetic
spin moment of the Fe atom is enhanced in comparison to the bulk fcc
Fe value of 2.2 $\mu$B. Second, in all three cases, after peaking at
about 2.61 $\mu$B for 3 substrate layers, the Fe atom takes on a
magnetic spin moment of about 2.56 $\mu$B, and the first W layer under
the Fe surface layer takes on a moment of about 0.1 $\mu$B in the
opposite direction (hence, it is antiferromagnetically coupled to the
Fe layer), consistent with the result of Ref.~\onlinecite{c:35}. The
remaining magnetic spin moments, including the interstitial moments,
are negligible (less than 1.5\%\ of the Fe moment).

Looking at the orbital moments in Tab.~\ref{t:3} ($\mu_L$), we observe
that while the spin moments do not change their sign with a change of
the magnetization axis, the orbital moments do. With magnetization
along the easy axis ({\bf M}$\|[1\bar{1}0]$), the orbital moment of
the Fe surface layer is very small, and in general couples
antiparallel to its spin moment. For the cases with 1--2 W substrate
layers, the orbital moment in the first substrate layer couples
antiparallel to its spin moment (and parallel to the spin moment of
Fe). However, already with the addition of the the third substrate
layer, the orbital moment of the first substrate layer couples in
parallel to its spin moment (and antiparallel to the Fe spin
moment). For magnetization along the hard axis ({\bf M}$\|[001]$) and
perpendicular to the surface plane ({\bf M}$\|[110]$), the coupling
picture is opposite. The orbital moment of the Fe atom is still very
small, but now coupled in parallel to its spin moment, and the orbital
moment of the first W substrate layer couples in parallel to its spin
moment for the cases with 1--2 W substrate layers (and antiparallel to
the spin moment of the Fe surface layer). The rather large orbital
moments of W$_1$ for $N_W=1$ and 2 might occur due to the fact that
these systems tend to have molecular properties rather than
solid-state ones. They might originate in a combination of the large
spin-orbit coupling found in W in combination with the large Fe
moment. From the addition of the third substrate layer and onwards,
the orbital moments of the first W atom couple antiparallel to their
spin moments (and parallel to the Fe spin moment). Thus, with respect
to the orbital moments, we conclude (i) that three substrate layers
already deliver a converged result, and (ii) that in agreement with
the observations made by Refs~\onlinecite{c:13,c:36}, Hund's 3rd rule
is broken for the W substrate.

For the perpendicular magnetization direction ({\bf M}$\|[110]$), our
results for $\mu_L$ differs qualitatively from those obtained in
Ref.~\onlinecite{c:13}. In order to explain this difference, a number
of calculations were performed on the symmetric slab, based on the
hypotheses that the difference is caused by (i) instabilities in the
computer code, (ii) the perturbative final addition of the spin-orbit
coupling in Ref.~\onlinecite{c:13}, (iii) the amount of ``vacuum''
between the slabs in the supercell calculation, or (iv) the breaking
of the symmetry in the slab (removing one of the Fe surfaces). A
repetition of the calculation in Ref.~\onlinecite{c:13} eliminates the
first hypothesis. A comparison between the repeated calculation and
one with spin-orbit coupling included selfconsistently leads to
elimination of the second hypothesis.\cite{c:35x} In the symmetric
slab, the Fe layers may couple electronically to each other either
through the W substrate or through the vacuum.  In
Ref.~\onlinecite{c:13}, the distance through the W substrate is 24.52
Bohr, and the distance through the vacuum is 26.01 Bohr.  Thus, in
order to test hypothesis (iii) against hypothesis (iv) we increase the
distance between the two Fe layers on the vacuum side to 76.54
Bohr.\cite{c:35y} The result of the calculation with the increased
spacing on the vacuum side between the two Fe surfaces is [in support
of hypothesis (iii)] an orbital moment for the Fe layer of
$0.00005\mu_B$ and an orbital moment for the first W layer of
$0.028\mu_B$, in agreement with Tab.~\ref{t:3}.  Thus, in agreement
with Refs~\onlinecite{c:36,c:37}, we may conclude from this little
excercise on the symmetric slabs and from Tab.~\ref{t:3} that both the
size as well as the alignment of the orbital moments in the Fe surface
layer and the first W substrate layer are not only very sensitive to
the local structure, but also to the direction of magnetization in the
Fe layer.

In order to determine the influence of the substrate layers on the
electronic properties of the magnetic surface, the evolution in the
density of states for the d-electrons (d-DOS) of the Fe atom has been
plotted in Fig.~\ref{f:2} (easy-axis magnetization, {\bf
M}$\|[1\bar{1}0]$), beginning with the Fe/W(110) bilayer result in
Fig.~\ref{f:2}(a). In Figs.~\ref{f:2}(b)--(h), the differences due to
the addition of a second, third, fourth, fifth, sixth, seventh, and
eighth substrate layer, respectively, are depicted on the same scale
as the d-DOS of the bilayer. It is clear from Fig.~\ref{f:2} that the
main features of the d-DOS of the Fe(110) surface layer are to be
found in a band between 4 eV above and 5 eV below the Fermi energy
($E_F=0$ in Fig.~\ref{f:2}). Also, we observe from Fig.~\ref{f:2} that
by adding substrate layers, the changes to the d-DOS of the Fe surface
layer become less and less pronounced, as the number of substrate
layers go up. Since already after the fourth or fifth substrate layer,
the d-DOS of the Fe surface layer has converged to within a few
percent of what it would be on bulk W(110), the d-DOS supports the
conclusion that the MAEs have reached their bulk value after adding
4-5 substrate layers.

When the d-DOS has converged, the differences in d-DOS due to changes
in the magnetization direction are very subtle. In order to illustrate
this, we have in Fig.~\ref{f:3}(a) plotted the d-DOS of an Fe(110)
monolayer with four W substrate layers underneath it for the three
cases where the magnetization direction is along one of the three main
crystallographic axes. From Fig.~\ref{f:3}(a) we notice that the d-DOS
is now almost identical in all three cases. In order to explore the
subtle differences in the d-DOS between the different magnetization
directions, we have in Figs.~\ref{f:3}(b)--(c) additionally plotted
the differences in the d-DOS between the easy axis, the hard axis, and
the axis perpendicular to the surface, i.e., the two quantities
d-DOS({\bf M}$\|[1\bar{1}0]$)$-$d-DOS({\bf M}$\|[001]$), and
d-DOS({\bf M}$\|[1\bar{1}0]$)$-$d-DOS({\bf M}$\|[110]$),
respectively. In contrast to the evolution of the d-DOS when adding
substrate layers, the differences in the d-DOS between the different
magnetization directions are too small to be plotted on the same scale
as the d-DOS itself. Since the MAEs are in the meV-range, this is to
be expected, since large changes in the d-DOS would lead to large
MAEs.

\section{Conclusions}\label{sec:IV}
We have shown that the magnetocrystalline anisotropy energy in an
Fe(110) monolayer on W(110) can be converged with respect to the
thickness of the substrate using {\em ab initio} methods. After
showing large changes for the first few substrate layers, it
stabilized close to 3 meV, close to the experimental value.  In
addition, the directions of the easy and hard axes came out
consistently, and in-plane. As expected from the large variations in
the MAEs with respect to the addition of the first few substrate
layers, also the density of the Fe d-states vary a lot during the
addition of the first few substrate layers, after which it stabilizes.
At the bulk-like substrate thicknesses, the differences between the
density of d-states between the different magnetization directions
reflects the fact that the MAE is in the meV regime (they are very
small).

\section*{Acknowledgements}
We acknowledge financial support from (i) the European Union FP5
Research Training Networks ``First-Principles Approach to the
Calculation of Optical Properties of Solids'' (EXCITING) and
``Dynamics in Magnetic Nanostructures'' (Dynamics), under Contracts
No. HPRN-CT-2002-00317 and HPRN-CT-2002-00289, respectively, (ii)
Deutsche Forschungsgemeinschaft SPP 1133 ``Ultraschnelle
Magnetisierungsprozesse'' and SPP 1153 ``Cluster in Kontakt mit
Oberfl{\"a}chen'', and (iii) Forschungsschwerpunkt des Landes
Rheinland-Pfalz ``Materialien f{\"u}r Mikro- und Nanosysteme" (MINAS).

\bibliographystyle{prsty} 

\begin{thebibliography}{10}

\bibitem{c:1} H.~J. Elmers, J. Hauschild, H. H{\"o}che, U. Gradmann,
H. Bethge, D. Heuer, and U. K{\"o}hler, Phys. Rev. Lett. {\bf 73}, 898
(1994).

\bibitem{c:2} R. R{\"o}hlsberger, J. Bansmann, V. Senz, K.~L. Jonas,
A. Bettac, U. Leopold, R. R{\"u}ffer, E. Burkel, and
K.~H. Meiwes-Broer, Phys. Rev. Lett. {\bf 86}, 5597 (2001).

\bibitem{c:3} U. Gradmann, M. Przybylski, H.~J. Elmers, and G. Liu,
Appl. Phys. A {\bf 49}, 563 (1989).

\bibitem{c:4} H.~J. Elmers and U. Gradmann, Appl. Phys. A {\bf 51},
255 (1990).

\bibitem{c:5} H.~J. Elmers, T. Furubayashi, M. Albrecht, and
U. Gradmann, J. Appl. Phys. {\bf 70}, 5764 (1991).

\bibitem{c:6} H.~J. Elmers, J. Hauschild, H. Fritzsche, G. Liu,
U. Gradmann, and U.  K{\"o}hler, Phys. Rev. Lett. {\bf 75}, 2031
(1995).

\bibitem{c:7} M. Bode, O. Pietzsch, A. Kubetzka, and R. Wiesendanger,
Phys. Rev. Lett. {\bf 92}, 067201 (2004).

\bibitem{c:8} S. Krause, L. Berbil-Bautista, M. Bode, and
R. Wiesendanger, Verhandl. DPG (VI) {\bf 41}, 1/MA27.10 (2006).

\bibitem{c:9} U. Gradmann, J. Korecki, and G. Waller, Appl. Phys. A
{\bf 39}, 101 (1986).

\bibitem{c:10} H.~J. Elmers, J. Hauschild, and U. Gradmann,
Phys. Rev. B {\bf 59}, 3688 (1999).

\bibitem{c:11} D. Sander, R. Skomski, C. Schmidthals, A. Enders, and
J. Kirschner, Phys. Rev.  Lett. {\bf 77}, 2566 (1996).

\bibitem{c:12} N. Weber, K. Wagner, H.~J. Elmers, J. Hauschild, and
U. Gradmann, Phys. Rev. B {\bf 55}, 14121 (1997).

\bibitem{c:13} X. Qian and W. H{\"u}bner, Phys. Rev. B {\bf 64},
092402 (2001).

\bibitem{c:14} M. Bode, R. Pascal, and R. Wiesendanger,
J. Vac. Sci. Technol. A {\bf 15}, 1285 (1997).

\bibitem{c:15} G. Waller and U. Gradmann, Phys. Rev. B {\bf 26}, 6330
(1982).

\bibitem{c:16} B. Hillebrands, P. Baumgart, and G. G{\"u}nterodt,
Phys. Rev. B {\bf 36}, 2450 (1987).

\bibitem{c:17} F. Gerhardter, Y. Li, and K. Baberschke, Phys. Rev. B
{\bf 47}, 11204 (1993).

\bibitem{c:18} I.-G. Baek, H.~G. Lee, H.-J. Kim, and E. Vescovo,
Phys. Rev. B {\bf 67}, 075401 (2003).

\bibitem{c:19} K. von Bergmann, M. Bode, and R. Wiesendanger,
Phys. Rev. B {\bf 70}, 174455 (2004).

\bibitem{c:20} A. Kubetzka, P. Ferriani, M. Bode, S. Heinze,
G. Bihlmayer, K. von Bergmann, O.  Pietzsch, S. Bl{\"u}gel, and
R. Wiesendanger, Phys. Rev. Lett. {\bf 94}, 087204 (2005).

\bibitem{c:21} D. Spi{\v{s}}{\'a}k and J. Hafner, Phys. Rev. B {\bf
70}, 195426 (2004).

\bibitem{c:22} P. Ferriani, S. Heinze, G. Bihlmayer, and
S. Bl{\"u}gel, Phys. Rev. B {\bf 72}, 024452 (2005).

\bibitem{c:23} X. Qian and W. H{\"u}bner, Phys. Rev. B {\bf 60}, 16192
(1999).

\bibitem{c:24} X. Qian, F. Wagner, M. Petersen, and W. H{\"u}bner,
J. Magn. Magn. Mater. {\bf 213}, 12 (2000).

\bibitem{c:25} I. Galanakis, M. Alouani, and H. Dreyss{\'e},
Phys. Rev. B {\bf 62}, 3923 (2000).

\bibitem{c:26} H.~L. Meyerheim, D. Sander, R. Popescu, J. Kirschner,
O. Robach, and S. Ferrer, Phys. Rev. Lett. {\bf 93}, 156105 (2004).

\bibitem{c:27} J.~W. Strutt, Phil. Trans. Roy. Soc. {\bf 161}, 77
(1870).

\bibitem{c:28} W. Ritz, J. reine angew. Math. {\bf 135}, 1 (1908).

\bibitem{c:29} E. Sj{\"o}stedt, L. Nordstr{\"o}m, and D.~J. Singh,
Solid State Commun. {\bf 114}, 15 (2000).

\bibitem{c:30} P. Blaha, K. Schwarz, G. K. H. Madsen, D. Kvasnicka, and
J. Luitz, {\em WIEN2k, An Augmented Plane Wave + Local Orbitals
Program for Calculating Crystal Properties} (Karlheinz Schwarz,
Vienna, 2001), ISBN 3-9501031-1-2.

\bibitem{c:31} D. Sander, A. Enders, and J. Kirschner, J. Magn.
Magn. Mater. {\bf 200}, 439 (1999).

\bibitem{c:32} J.~P. Perdew, K. Burke, and M. Ernzerhof, Phys. Rev.
Lett. {\bf 77}, 3865 (1996); {\em ibid} {\bf 78}, 1396(E) (1997).

\bibitem{c:32a} It should be noted here, that in order to secure a
sufficiently accurate {\em magnetic} ground state, the spin-orbit
coupling should be included self-consistently in the variational
minimization of the total energy. Previously,\cite{c:23}, the {\em
nonmagnetic} ground state was converged self-consistently, and the
spin-orbit coupling was added perturbatively at the end of the
calculation.

\bibitem{c:32b} The core states were calculated with 781 radial mesh
points. 

\bibitem{c:33} M. Pratzer, H.~J. Elmers, M. Bode, O. Pietzsch,
A. Kubetzka, and R.  Wiesendanger, Phys. Rev. Lett. {\bf 87}, 127201
(2001).

\bibitem{c:33a} H.~J. Elmers, G. Liu, and U. Gradmann, Phys. Rev.
Lett. {\bf 63}, 566 (1989).

\bibitem{c:33b} J. Hauschild, H.~J. Elmers, and U. Gradmann, Phys.
Rev. B {\bf 57}, 677 (1998).

\bibitem{c:34} With respect to the results of the symmetric Fe/W(110)
structures of Ref.~\onlinecite{c:13}, it is interesting to notice that
removing the Fe layer on one side of the W(110) slab influences the
MAEs in a manner a bit similar to (although not as strong as) that of
adding an extra Fe layer on both sides of the slab, i.e., we obtain an
in-plane hard axis and a slightly smaller anisotropy energy.

\bibitem{c:35} X. Qian and W. H{\"u}bner, Phys. Rev. B {\bf 67},
184414 (2003).

\bibitem{c:35x} The results for the orbital moments in these two
calculations are consistent to within 10\%\ in magnitude with those in
Ref.~\onlinecite{c:35}, and the signs are identical. In order to
assure sufficient accuracy in these two calculations, we have used
$T_c=9$ and \#{\bf k}$=2883$ here.

\bibitem{c:35y} Thereby making the height of the unit cell comparable
to the one used in the present calculations. For this comparison
calculation, we used $T_c=9$ and \#{\bf k}$=1521$.

\bibitem{c:36} F. Wilhelm, P. Poulopoulos, H. Wende, A. Scherz,
K. Baberschke, M. Angelakeris, N.~K. Flevaris, and A. Rogalev,
Phys. Rev. Lett. {\bf 87}, 207202 (2001); {\em ibid} {\bf 90}, 129702
(2003).

\bibitem{c:37} R. Tyer, G. van der Laan, W. M. Temmermann, and
Z. Szotek, Phys. Rev. Lett. {\bf 90}, 129701 (2003).

\end{thebibliography}

\end{document}